\documentclass[preprint]{aastex7}

\newcommand{\gsim}{\mbox{\hspace{.2em}\raisebox{.5ex}{$>$}\hspace{-.8em}\raisebox{-.5ex}{$\sim$}\hspace{.2em}}}
\newcommand{\lsim}{\mbox{\hspace{.2em}\raisebox{.5ex}{$<$}\hspace{-.8em}\raisebox{-.5ex}{$\sim$}\hspace{.2em}}}

\newcommand{\E}[1]{\times 10^{#1}}
\newcommand{\twCO}{$^{12}$CO}  \newcommand{\thCO}{$^{13}$CO}

\newcommand{\HI}{\mbox{H\,\textsc{i}}}

\newcommand{\s}{\,{\rm s}}      \newcommand{\ps}{\,{\rm s}^{-1}}
\newcommand{\yr}{\,{\rm yr}}    \newcommand{\Msun}{M_{\odot}}   
    \newcommand{\km}{\,{\rm km}}
\newcommand{\kpc}{\,{\rm kpc}}
        \newcommand{\K}{\,{\rm K}}
    
    \newcommand{\G}{\,{\rm G}}

\newcommand{\VLSR}{V_{\rm LSR}}

\begin{document}

\title{
How Is Cold Gas Loaded into Galactic Nuclear Outflows?
}

\shorttitle{Cold Gas Loading from the Warped Inner Disk into Nuclear Outflows} 

\correspondingauthor{Yang Su}
\email{yangsu@pmo.ac.cn}

\author[orcid=0000-0002-0197-470X,sname='Su']{Yang Su}
\affil{Purple Mountain Observatory and Key Laboratory of Radio Astronomy, 
Chinese Academy of Sciences, 10 Yuanhua Road, Nanjing 210023, China}
\affiliation{School of Astronomy and Space Science, University of Science and
Technology of China, 96 Jinzhai Road, Hefei 230026, China}
\email{yangsu@pmo.ac.cn}

\author[orcid=0000-0002-0409-7466,sname='Liu']{Xin Liu}
\affil{Purple Mountain Observatory and Key Laboratory of Radio Astronomy,
Chinese Academy of Sciences, 10 Yuanhua Road, Nanjing 210023, China}
\affiliation{School of Astronomy and Space Science, University of Science and
Technology of China, 96 Jinzhai Road, Hefei 230026, China}
\email{liuxin@pmo.ac.cn}

\author[orcid=0009-0002-2379-4395,sname='Zhang']{Shiyu Zhang}
\affiliation{Purple Mountain Observatory and Key Laboratory of Radio Astronomy, 
Chinese Academy of Sciences, 10 Yuanhua Road, Nanjing 210023, China}
\email{syzhang@pmo.ac.cn}

\author[orcid=0000-0001-7768-7320,sname='Yang']{Ji Yang}
\affiliation{Purple Mountain Observatory and Key Laboratory of Radio Astronomy,
Chinese Academy of Sciences, 10 Yuanhua Road, Nanjing 210023, China}
\affiliation{School of Astronomy and Space Science, University of Science and
Technology of China, 96 Jinzhai Road, Hefei 230026, China}
\email{jiyang@pmo.ac.cn}

\author[orcid=0000-0002-3904-1622,sname='Sun']{Yan Sun}
\affiliation{Purple Mountain Observatory and Key Laboratory of Radio Astronomy, 
Chinese Academy of Sciences, 10 Yuanhua Road, Nanjing 210023, China}
\affiliation{School of Astronomy and Space Science, University of Science and
Technology of China, 96 Jinzhai Road, Hefei 230026, China}
\email{yansun@pmo.ac.cn}

\author[orcid=0000-0003-2549-7247,sname='Zhang']{Shaobo Zhang}
\affiliation{Purple Mountain Observatory and Key Laboratory of Radio Astronomy, 
Chinese Academy of Sciences, 10 Yuanhua Road, Nanjing 210023, China}
\email{shbzhang@pmo.ac.cn}

\author[orcid=0000-0002-7489-0179,sname='Du']{Fujun Du}
\affiliation{Purple Mountain Observatory and Key Laboratory of Radio Astronomy,
Chinese Academy of Sciences, 10 Yuanhua Road, Nanjing 210023, China}
\affiliation{School of Astronomy and Space Science, University of Science and
Technology of China, 96 Jinzhai Road, Hefei 230026, China}
\email{fjdu@pmo.ac.cn}

\author[orcid=0000-0003-2418-3350,sname='Zhou']{Xin Zhou}
\affiliation{Purple Mountain Observatory and Key Laboratory of Radio Astronomy,
Chinese Academy of Sciences, 10 Yuanhua Road, Nanjing 210023, China}
\email{xinzhou@pmo.ac.cn}

\author[orcid=0000-0003-4586-7751,sname='Yan']{Qing-Zeng Yan}
\affiliation{Purple Mountain Observatory and Key Laboratory of Radio Astronomy,
Chinese Academy of Sciences, 10 Yuanhua Road, Nanjing 210023, China}
\email{qzyan@pmo.ac.cn}

\author[orcid=0000-0003-3151-8964,sname='Chen']{Xuepeng Chen}
\affiliation{Purple Mountain Observatory and Key Laboratory of Radio Astronomy, 
Chinese Academy of Sciences, 10 Yuanhua Road, Nanjing 210023, China}
\affiliation{School of Astronomy and Space Science, University of Science and
Technology of China, 96 Jinzhai Road, Hefei 230026, China}
\email{xpchen@pmo.ac.cn}

\begin{abstract}
The origin of the multiphase gas within the Fermi/eROSITA bubbles is crucial 
for understanding Galactic center (GC) feedback. We use HI4PI data to investigate 
the kinematics and physical properties of high-velocity clouds (HVCs) toward the GC 
region ($l=+25^{\circ}$ to $-10^{\circ}$). Our results reveal that the HVCs exhibit a 
distinct asymmetric distribution, closely associated with bar-driven tilted dust 
lanes and distorted overshooting streams. We propose that powerful nuclear 
outflows interact with these gas-rich, off-plane structures, striping and entraining 
cold gas from the outer Galactic regions ($R_{\rm GC}\sim$~0.5--1.7~kpc) rather than 
solely from the central molecular zone (CMZ; $R_{\rm GC}\lsim$~0.3~kpc). 
In this scenario, as the Galactic bar drives gas inflows along the dust lanes, nuclear 
outflows simultaneously break through the CMZ, sweeping up and ablating cold gas from 
the boundary layer of these preexisting structures. This process naturally accounts 
for the observed high turbulence, complex spectral signatures, and anomalous 
spatial-kinematic gas patterns, as well as multiwavelength asymmetries of the bubbles. 
The HVCs are accelerated to about 230--340~km~s$^{-1}$ over a dynamical time of 
$\sim$3--6~Myr. When the multiphase, inhomogeneous composition of the gas is included, 
the estimated gas outflow rate in on the order of $\sim1~\Msun\yr^{-1}$. This value 
is comparable to the bar-driven inflow rate, indicating a tightly coupled gas cycle 
in the inner Galaxy. 
Our research highlights the critical role of bar-driven gas dynamics and nuclear 
feedback in the secular evolution of the Milky Way, offering a valuable paradigm for 
investigating the gas outflow--inflow cycle in external galaxies.
\end{abstract}


\section{Introduction}
The dynamics and physical properties of gas reservoirs are tightly linked to 
galactic evolution and the baryon cycle. As an important component of galaxies, 
cold gas acts not only as the primary fuel for star formation but also plays a 
fundamental role in shaping large-scale galactic structures (disks, spiral arms, 
and bar-driven inflows and outflows dominated by nuclear winds, for example). 
The interplay between cold and multiphase gas, together with the global 
interconnection of baryonic matter, serves as a crucial role in regulating 
the lifecycle of galaxies
\citep[e.g.,][]{2022ARA&A..60..319S,2023ARA&A..61...19M,2024ARA&A..62..369S}.

The Milky Way provides a unique observational 
platform for resolving its detailed structure and refining our understanding 
of galactic evolution and history. Generally, the Galactic bar drives radial 
gas inflows from the Galactic plane (GP) into the central molecular zone 
\citep[CMZ; e.g.,][]{1992MNRAS.259..345A,1999A&A...345..787F,2008A&A...489..115R,
2019MNRAS.484.1213S,2023ASPC..534...83H,2024ApJ...971L...6S}. 
Simultaneously, starburst episodes in the CMZ and/or accretion activity of the 
supermassive black hole (SMBH) in the Galactic center (GC) generate powerful nuclear 
winds. The winds expel gas vertically, giving rise to large-scale symmetrical lobes 
that extend above and below the GP \citep[the Fermi/eROSITA bubbles;][]
{2010ApJ...724.1044S,2016ApJ...829....9M,2019Natur.573..235H,2020Natur.588..227P}. 
This dynamical coupling of gas inflows, star formation and SMBH feedback, and 
Galactic outflows forms a self-regulating baryonic cycle in the Milky Way 
\citep[][]{2020A&ARv..28....2V,2021NewAR..9301630B,2024A&ARv..32....1S}. 
The intricate dynamical features observed in our Galaxy offer an
excellent laboratory for investigating the complex ecosystems of galaxies 
beyond our own.

Based on the CO data from the Milky Way Imaging Scroll Painting \citep[MWISP;][]
{2019ApJS..240....9S}, we show that the Galactic bar plays a critical role in
governing the evolution of tilted and distorted gas streams by
(1) facilitating radial gas transport from the Galactic disk at Galactocentric 
distances of $R_{\rm GC}\sim$~3.0--3.4~kpc to the CMZ at $R_{\rm GC}\sim$~0.3~kpc,
(2) redistributing angular momentum through collisional dissipation in cold gas
flows and direct bar-gas interactions,
and (3) regulating gas inflow to the CMZ via the overshoot 
mechanism \citep[e.g.,][]{2024ApJ...971L...6S,2025ApJ...984..109S}. 
Only $\sim$30\% of the bar-driven gas directly enter the CMZ, while the
remaining gas undergoes several orbital periods 
(e.g., tens of millions of years) before finally settling into the CMZ. 
This self-consistent picture naturally reproduces the observed spatial and
kinematic properties of molecular gas within the inner GP.

On the other hand, while significant progress has been made over the past 
decade in studying Galactic nuclear outflows, a comprehensive understanding 
of its impact on gas distribution and kinematic features remains limited. 
How is cold gas efficiently entrained into the halo by a hot, high-velocity wind? 
How do molecular gas and dust survive or reform under such extreme conditions?
What is the outflow rate driven by the nuclear wind? And how does the coupling 
between bar-driven inflows and nuclear-wind-dominated outflows collectively govern
the large-scale structure and long-term evolution of the Milky Way?
These important questions are still subject to intense debate because of 
observational challenges and the inherent complexities of modeling multiphase 
gas dynamics
\citep[e.g.,][]{2015ApJ...808..107C,2017ApJ...834..191B,2017MNRAS.466.1213K}.

In this work, we investigate how Galactic nuclear outflows strip cold gas to 
supply the Fermi/eROSITA bubbles, an interaction that accounts for the entrainment 
of multiphase gas from off-plane, gas-rich structures. In Section 2, we illustrate 
the close connection between the high-velocity cloud (HVCs) outside the GP 
and the bar-aligned inflows toward the CMZ. In Section 3, we argue that the
outflows interact with preexisting off-plane clouds and largely enhance gas 
loading through the entrainment of multiphase gas from tilted and warped gas 
structures. Section 4 presents a unified framework of gas circulation in the 
inner 3~kpc of the Milky Way, where the tilted and warped gas structures 
become the key dynamical 
link connecting bar-driven inflows with nuclear-wind-dominated outflows.

\section{Tilted High-velocity Clouds off the Galactic Disk}
Early studies of GC gas dynamics, dating back to 1980s \HI\ and CO observations
\citep[e.g.,][]{1978ApJ...225..815B,1978ApJ...226..790L,1980ApJ...236..779L}, 
initially identified structures exhibiting kinematic anomalies inconsistent
with an axisymmetric distribution and circular motion. These discoveries attracted 
limited attention until seminal studies in the 1990s began integrating gas 
dynamics into models that included a central bar potential in the inner Galaxy
\citep[e.g.,][]{1991MNRAS.252..210B,1991ApJ...379..631B}.

Recent studies have resolved two phenomena toward the central 3~kpc 
of the Milky Way:
(1) global CO inflows driven by the Galactic bar along the near dust lane, 
exhibiting kinematics consistent with X1 orbit streaming motions,
and (2) coherent and extended CO streams that overshoot from the far dust 
lane ($l\lsim+0^{\circ}$) to remarkable heights of z$\lsim-170$~pc 
\citep[][]{2024ApJ...971L...6S,2025ApJ...984..109S}. These findings provide 
detailed insight into the gravitational influence of the Galactic bar on gas 
streams (e.g., highly elongated and distorted structures, localized velocity 
gradients and shears, and hierarchically fragmented configurations, etc.).
It is noteworthy that these two identified large-scale 
gas structures are located predominantly below the GP 
(i.e., at negative Galactic latitudes for $l\gsim+1^{\circ}$).

We utilize HI4PI data \citep[$16'$ resolution and 43~mK sensitivity; see the 
HI4PI survey;][]{2016A&A...594A.116H} to investigate the distribution and 
properties of atomic gas toward the GC region
($l=[+25\fdg0, -10\fdg0]$, $|b|\lsim15\fdg0$). This analysis is
supplemented with MWISP data \citep[][]{2019ApJS..240....9S}, 
which provide higher spatial resolution ($50''$) at a sensitivity of $\sim0.5$~K
($\sim0.3$~K) to examine the associated molecular structures traced by \twCO\ 
(\thCO) emission.
The CO observations are limited to a small portion of the \HI\
field (i.e., the region of $l=[+9\fdg7, +1\fdg2]$, $|b|\lsim2\fdg5$ and
$l\gsim10\fdg0$, $|b|\lsim5\fdg5-7\fdg5$).
The phase-I dataset of the MWISP survey has been released
\citep[MWISP Data Release 1; https://doi.org/10.57760/sciencedb.27351; 
see details in][]{2025arXiv251208260Y}.

Using an algorithmic pipeline \citep[GDCluster;][]{2025ApJ...994...76L},
we extracted atomic clouds with extreme velocities to investigate the
distribution and physical properties of HVCs toward the GC. 
Beginning with \HI\ spectra represented by multiple Gaussian components, 
GDCluster employs a derivative spectroscopy technique to estimate the 
initial number of components and their preliminary parameters (i.e., 
amplitude, center velocity, width, etc.). Assuming localized uniformity 
of \HI\ clouds, GDCluster refines these parameters by incorporating data 
from adjacent sight lines, reducing solution degeneracies through spatial 
regularization. Following least-squares fitting, all Gaussian 
components are grouped into physically coherent structures based on their 
spatial and kinematic similarities. In the implementation, kinematic 
components are resolved as distinct structures when the velocity separation 
between Gaussian peaks meets the Rayleigh criterion ($\geq1.22\times$FWHM) 
and the peak intensities exceed 3 times the threshold 
(i.e., $\gsim$0.13~K for the \HI\ data).

\begin{figure}
\vspace{-5ex}
\gridline{\hspace{-12ex} \fig{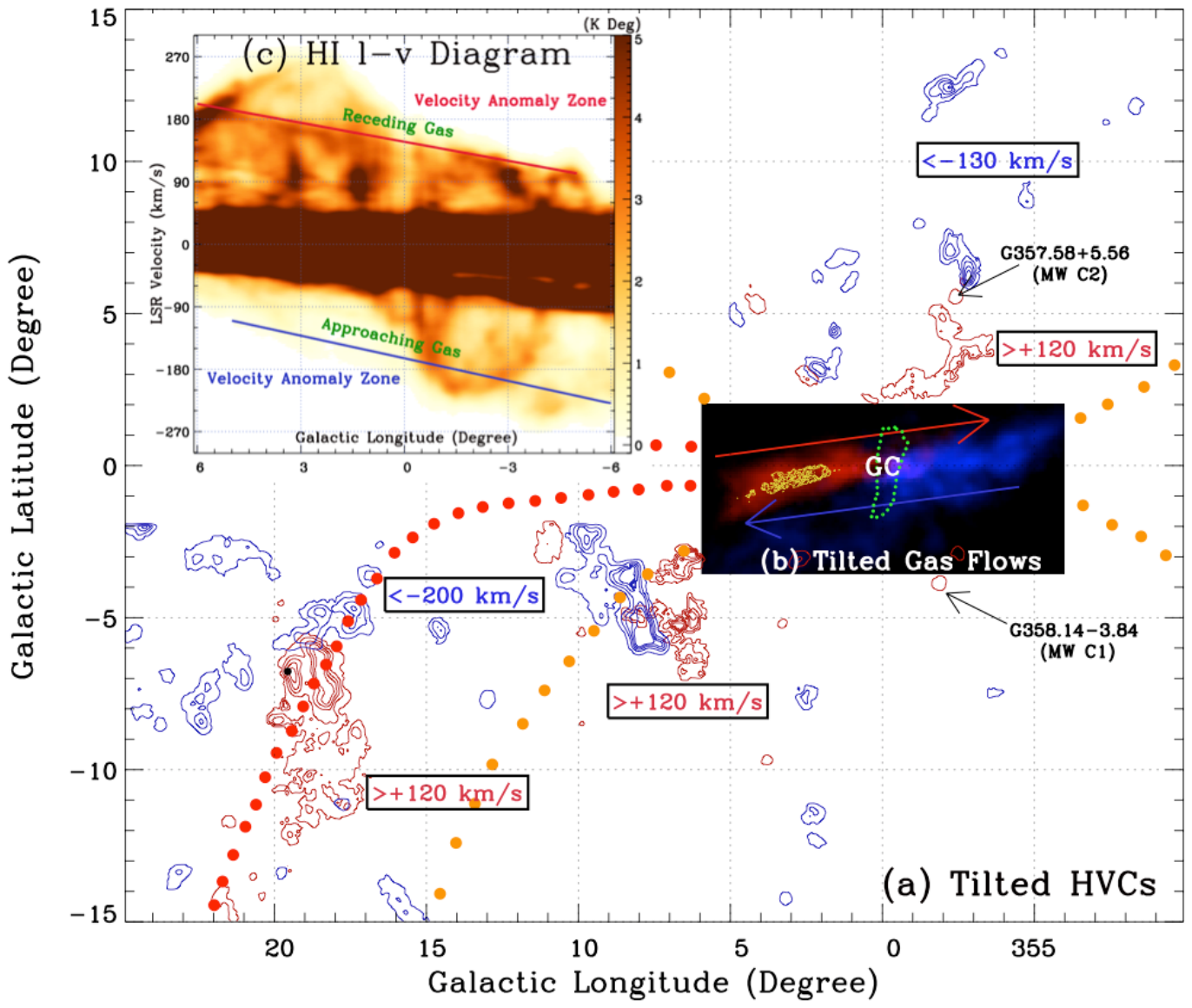}{1.2\textwidth}{} 
         }
\vspace{-9ex}
\caption{
Panel (a) illustrates the large-scale tilt of off-plane HVCs toward the GC.
The contours show the intensity of the HVC, and the colors represent its
extreme velocities at $|b|\gsim2\fdg0$. The blue color traces
the approaching HVCs (starting from 1~K~km~s$^{-1}$ with a step of 4~K~km~s$^{-1}$),
while the red is for receding HVCs (starting from 2~K~km~s$^{-1}$ with a step of 
12~K~km~s$^{-1}$). Two HVCs, labeled MW C1 and MW C2 
\citep[see][]{2020Natur.584..364D}, are marked in the figure.
Note that the extreme negative-velocity HVCs in the range $l\lsim0^{\circ}$ and
$b=-3\fdg5$ to $-6\fdg5$ are not shown due to potential 
contamination of approaching gas from the far dust lane.
The black point marks the MWISP-detected CO cloud at ($l=19\fdg56, b=-6\fdg77$).
Panel (b) displays the tilted gas flows driven by the Galactic bar within 
$l=[+6^{\circ},-6^{\circ}]$ and $b=[-3\fdg5, +2^{\circ}]$. 
Along the tilted dust lanes, the streams flow towards the CMZ. Red indicates 
receding atomic gas from the near dust lane, while blue shows approaching gas
from the far dust lane. Superimposed yellow contours are molecular inflows from
MWISP CO data \citep[starting from 20~K~km~s$^{-1}$ and increase with a step of 
30~K~km~s$^{-1}$; see ballisticlike CO streams in][]{2024ApJ...971L...6S}. 
Panel (c) presents the longitude-velocity diagram of \HI\ gas, highlighting 
anomalous velocity zones across different Galactic longitude regions.
The red and blue lines mark the lower and upper velocity integration limits 
for the receding and approaching gas structures identified in 
panel (b), respectively. 
Nearly all the high-latitude HVCs that we have identified fall within 
these anomalous spatial and velocity ranges (Figure~\ref{fig:f2}).
The orange and red dots denote the morphology of the Fermi bubbles 
(the interior of hot, shocked wind) and the associated shell structures
\citep[the shocked ISM and CGM material; see the GeV gamma-ray and the corresponding
thermal gas from the X-ray observations;][]{2010ApJ...724.1044S,2016ApJ...829....9M},
respectively.
The green dots mark the 430~pc bipolar radio bubbles, seen in the MeerKAT
1.284~GHz map toward the GC \citep[][]{2019Natur.573..235H}.
\label{fig:f1}}
\end{figure}

Building on a global kinematic perspective of the identified HVC samples, 
we report the discovery of a tilted distribution of extreme-velocity atomic gas
relative to the GP. 
Figure~\ref{fig:f1}a shows an antisymmetric distribution of HVCs. 
At $l\gsim0^{\circ}$, HVCs are predominantly located at $b\lsim-2^{\circ}$, 
while the opposite trend is seen at $l\lsim0^{\circ}$ and $b\gsim+2^{\circ}$.
HVCs at $l\gsim7^{\circ}$ even exhibit extreme negative local standard of rest
(LSR) velocities as low as $-200$~km~s$^{-1}$ to $-300$~km~s$^{-1}$, 
whereas their counterparts at $l\sim-2\fdg5$ display positive velocities 
reaching up to $+270$~km~s$^{-1}$ 
\citep[e.g., G357.68$+$5.56 or MW~C2 in Figure~\ref{fig:f1}a; also see][]
{2020Natur.584..364D}.

The kinematically anomalous HVCs show spatial correlation with the
tilted dust lanes, where bar-driven gas inflows are located 
\citep[for details delineated by CO emission, see]
[]{2024ApJ...971L...6S,2025ApJ...984..109S}. 
Interestingly, the HVCs with the most extreme negative velocities are 
situated adjacent to vertically extended gas structures that have positive
velocities (e.g., red contours in $l\gsim+5^{\circ}$ of Figure~\ref{fig:f1}a).
The close spatial relationship likely suggests a common origin for the two
kinematically distinct populations (Section 3).
The MWISP survey covers a portion of the region of interest and detects the 
presence of a molecular cloud located at ($l=19\fdg56, b=-6\fdg77$) with 
$T_{\rm {peak}}\gsim0.5$K, $\VLSR\sim120$~km~s$^{-1}$, and size$\sim$~1.5~arcminutes.
The peak CO intensity is approximately one-tenth of the corresponding \HI\ 
intensity, indicating that deeper observations could detect 
fainter molecular gas emission associated with similar HVCs.

The inclination angle between the bar major axis and the line connecting 
the Sun and GC is estimated 
to be $\phi_{\rm bar}=23^{\circ}\pm3^{\circ}$ \citep[][]{2024ApJ...971L...6S}. 
Here, $\phi_{\rm bar}=23^{\circ}$ is derived from a simple geometry and does 
not account for the bending of gas streams during bar-driven accretion. 
From a simple geometric model at the point of maximum bending near the half-bar region 
(i.e., $l=6^{\circ}-7^{\circ}$ or $R_{\rm GC}\sim1.5-1.7$~kpc), the inclination of the 
leading inflows is derived to be $\sim34^{\circ}$, a value roughly 1.5 times greater 
than $\phi_{\rm bar}$.
At $R_{\rm GC}\sim1$~kpc ($l\sim5\fdg5$), the corresponding inclination angle 
approaches $\sim2\times\phi_{\rm bar}\sim46^{\circ}$. The larger inclination 
angle agrees with previous results \citep[e.g., $\sim40^{\circ}-50^{\circ}$ in]
[]{1980ApJ...236..779L}, denoting the mild bending of gas flows within the 
region of $R_{\rm GC}\lsim1.5-1.7$~kpc. 

Figure~\ref{fig:f1}b shows the tilted \HI\ gas near the GP,
demonstrating both receding (red) and approaching (blue) flows as
identified in the $l-v$ diagram (see Figure~\ref{fig:f1}c).
According to the \HI\ distribution from $l=-6^{\circ}$ to $+6^{\circ}$ and
$b=-3\fdg5$ to $+2^{\circ}$, the tilt angle of atomic inflows is estimated
to be $\theta_{\rm gas\ flows}\sim15^{\circ}$ by adopting an inclination angle
of 2$\times\phi_{\rm bar}$ in the inner 1~kpc region. The tilted feature relative 
to the GP aligns closely with the geometry of the CO inflow, where the molecular 
gas stream follows a ballisticlike trajectory toward the GC
\citep[yellow contours in Figure~\ref{fig:f1}b; also see][]{2024ApJ...971L...6S}.
For the anomalous HVCs at $l\sim+8^{\circ}$ and $b\sim-5^{\circ}$, the maximum
tilt angle is $\sim20^{\circ}$ when accounting for the inclination angle of
$\sim1.5\times\phi_{\rm bar}$. 
Within the bar- and bulge-dominated inner disk, the most severe gas warp occurs 
near the midpoint of the gas inflows at $R_{\rm GC}\sim1.5-1.7$~kpc.

The transition of gas flows from x1 to x2 orbits is likely a gradual process governed 
by the gravitational potential of the bar and multiple physical mechanisms, during which 
the inflow inclination evolves from $\sim23^{\circ}$ to $\sim113^{\circ}$ as gas moves 
from the GP into the CMZ. This dynamics imprints a large-scale warp on the 
gas disk within the inner 1.5--1.7~kpc, which naturally explains the more pronounced 
left-side ($l\gsim0^{\circ}, b\lsim0^{\circ}$) extension due to its proximity to the Sun. 
Simultaneously, in the outer bar and bulge region where gravitational confinement is 
weaker, overshooting gas streams become more prominently distorted, exhibiting negative 
velocities at $l\gsim0^{\circ}$, $b\lsim0^{\circ}$ and positive velocities at 
$l\lsim0^{\circ}$, $b\lsim0^{\circ}$. 
The distorted gas structures and their kinematic patterns perpendicular to the GP 
have coevolved and are fundamentally linked to the warped gaseous disk in the 
Milky Way's inner region.

Additionally, a substantial fraction of ionized gas coexists with neutral 
hydrogen gas \citep[][]{2017ApJS..232...25S,
2020ApJ...899L..11K,2020SciA....6.9711K,2021ApJ...923L..11C}, 
indicating the warped \HI\ structure acts as an important reservoir for 
the surrounding ionized gas. 
Collectively, the tilted CO dust lanes and overshooting streams, as well as the 
warped \HI\ disk and associated ionized gas, organize into a coherent yet deformed 
multiphase gas structure in the inner Galaxy at $R_{\rm GC}\sim0.5-1.7$~kpc. 
As a result, considerable multiphase gas is capable of reaching several 
hundred parsecs beyond the Galactic disk. This preexisting extraplanar gas 
acts as seeds for the HVCs that are entrained by Galactic
outflows (Section 3).

\section{Gone with the Wind}
Extensive observations and simulations have been conducted on the Galactic 
nuclear winds and outflows and their associated physical processes \citep[e.g.,][]
{2020A&ARv..28....2V,2024A&ARv..32....1S}. Surprisingly, however, little attention 
has been paid to connecting the macroscopic characteristics of the Fermi/eROSITA 
bubbles with the tilted and warped gas structure at $R_{\rm GC}\lsim1.7$~kpc. 
In this section, we propose that the anomalous HVCs toward the GC mainly 
originate from interactions between the Galactic outflows and these extraplanar 
structures (i.e., tilted dust lanes and distorted overshooting streams that are 
situated hundreds of parsecs away from the GP). 

\begin{figure}
\vspace{-5ex}
\gridline{
  \hspace{-.5ex} \fig{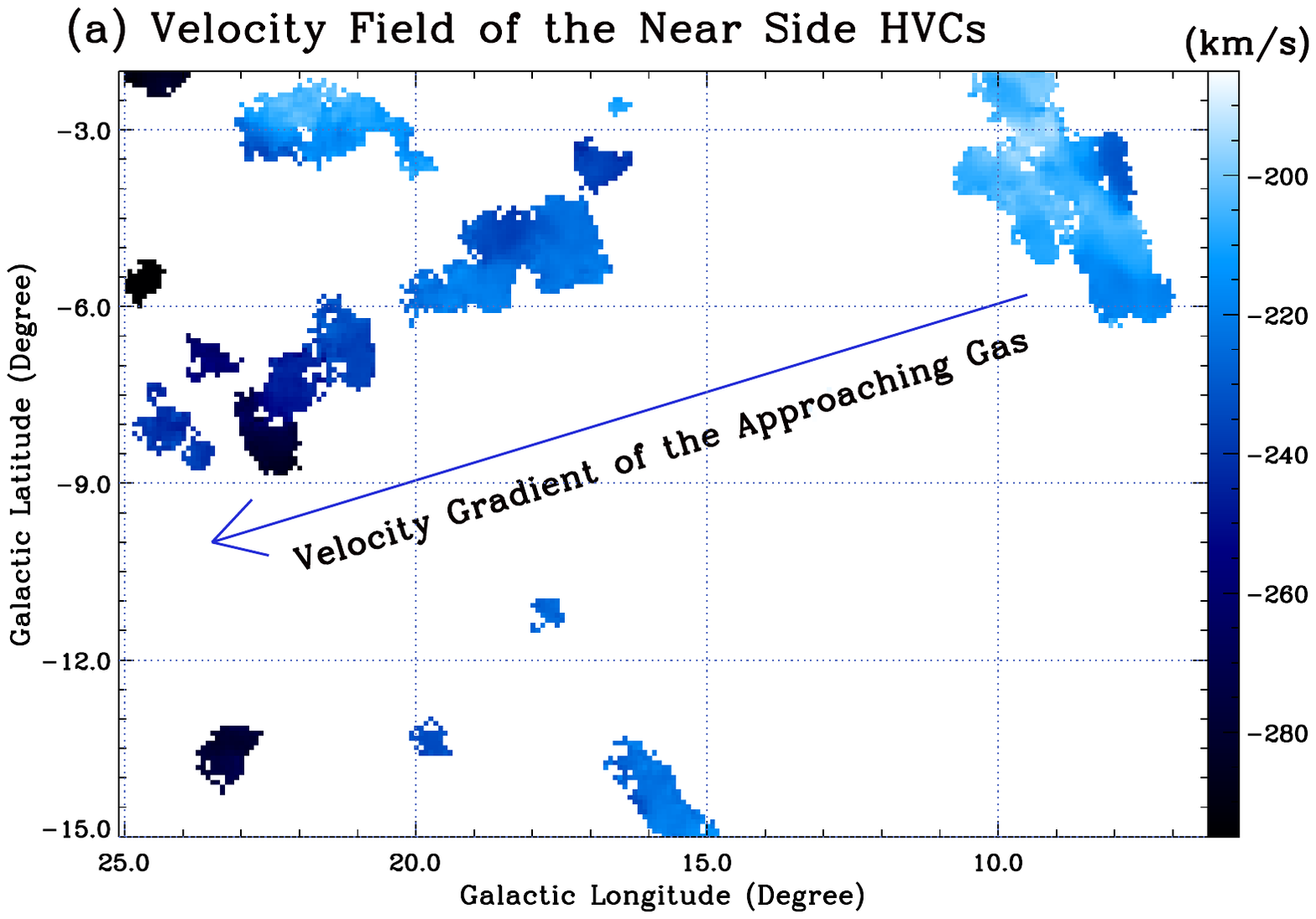}{0.85\textwidth}{} 
         }
\vspace{-12ex}
\gridline{
  \hspace{-1.5ex} \fig{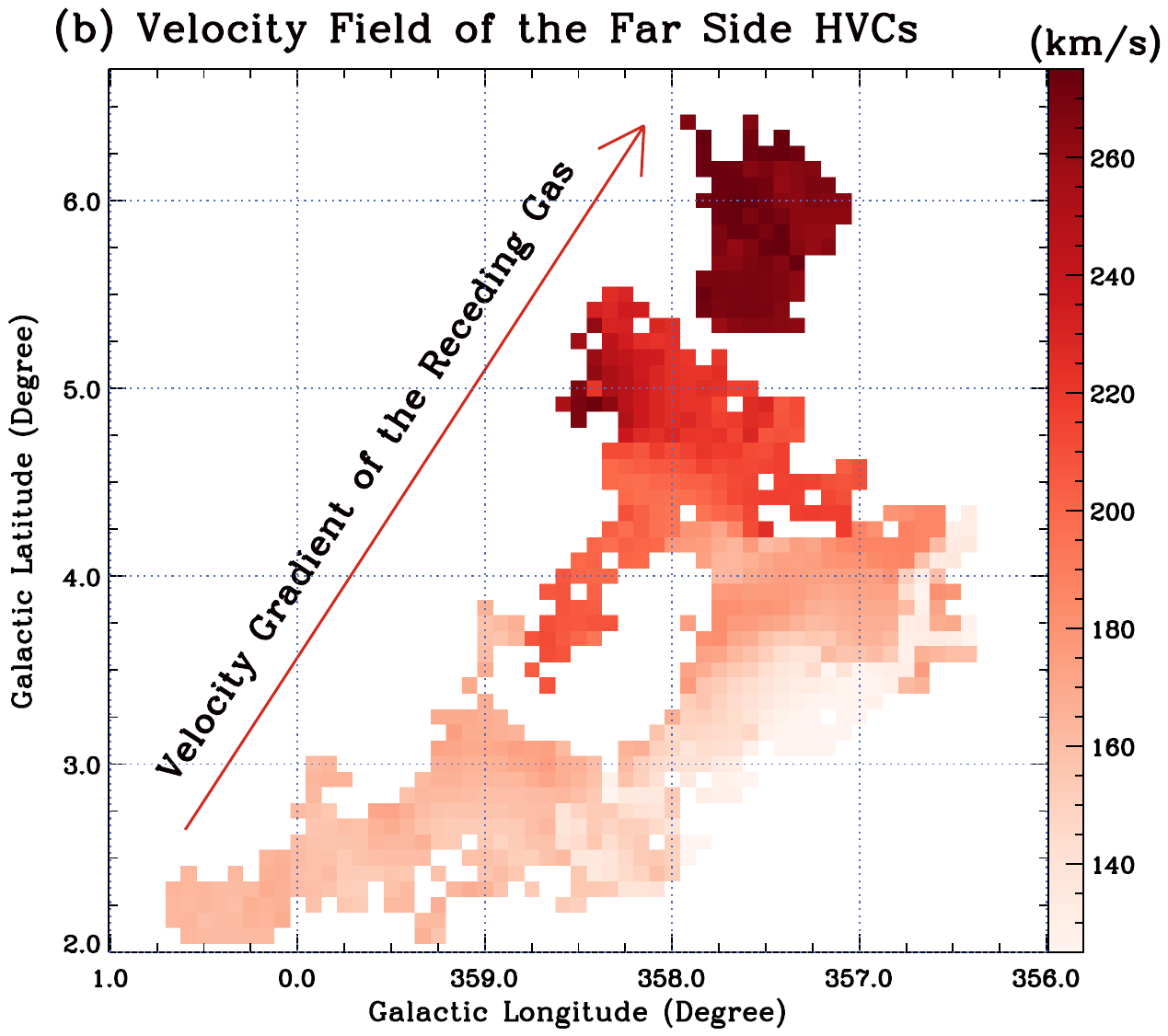}{0.95\textwidth}{} 
         }
\vspace{-8ex}
\caption{
Panel (a) displays the velocity distribution of extreme negative-velocity gas in the 
near-side region, where the blue arrow indicates progressively decreasing velocities 
(accelerating approach) with increasing $l$ and decreasing $b$. 
Panel (b) shows the velocity field of extreme positive-velocity gas in the far-side 
region, with the red arrow marking gradually increasing velocities (accelerating recession) 
as $l$ decreases and $b$ increases.
\label{fig:f2}}
\end{figure}

Anomalous HVCs are predominantly located near the tilted gas lanes, 
indicating a spatial association between them (Figure~\ref{fig:f1}).
As shown in Figure~\ref{fig:f2}, the HVCs toward $l\lsim~0^{\circ}$ and 
$b\gsim~+2^{\circ}$ exhibit a highly spatially coherent velocity gradient, 
where the LSR velocity increases from $\sim+120$~km~s$^{-1}$ at $b\sim+2^{\circ}$ 
to $\sim+270$~km~s$^{-1}$ at $b\sim+6^{\circ}$. This can be interpreted 
as evidence for an accelerating outflow originating from the far side of the 
Galaxy at a distance $\gsim$9~kpc. If these HVCs originated within 8~kpc, 
they would require significant deceleration from higher $b$ to the GP. 
This is inconsistent with the observed data and the established kinematic 
framework of the region \citep[i.e., coherent gas in $l\lsim0^{\circ}, b
\gsim2^{\circ}$, and the corresponding $\VLSR\gsim+120$~km~s$^{-1}$; 
see Figure~3 in][]{2025ApJ...984..109S}. 

In the region of $l\lsim~0^{\circ}$ and $b\gsim~+2^{\circ}$, the approaching 
gas from the far dust lane indeed shows deceleration, with velocity increasing 
from $\sim-160$~km~s$^{-1}$ at $b=+$6\fdg7 to $\sim-140$~km~s$^{-1}$ at 
$b=+$5\fdg9 (see blue contours in Figure~\ref{fig:f1}). The self-consistent 
interpretation likewise applies to the HVCs in $l\gsim~+5^{\circ}$ and 
$b\lsim~-2^{\circ}$ region, i.e., the gas traced by the blue contours is 
accelerating toward us, while the gas with the red contours exhibit behavior 
consistent with deceleration at a distance $\lsim$7~kpc.

The systematic outward velocity gradient observed in these anomalous HVCs 
suggests radial acceleration and ongoing dissipation with increasing $R_{\rm GC}$. 
This kinematic signature is distinct from bar-driven inflow patterns
\citep[Figure~\ref{fig:f1}c; also see][]{2025ApJ...984..109S}. 
The HVCs frequently exhibit cometary structures, with their heads facing the GC and 
tails extending outward, or they form arcs with their apexes point toward the GC.
Notably, the HVC morphology in the near region (e.g., $l\sim~+9^{\circ}$ 
and $b\sim~-4^{\circ}$) differs substantially from overshooting streams that fall back 
to the GP \citep[see Figure~2c in][]{2025ApJ...984..109S}. Similarly, on the far 
side (e.g., $l\lsim~0^{\circ}$ and $b\gsim~+2^{\circ}$), the extreme positive-velocity 
HVCs are also kinematically distinct from the twisted, disk-returning inflows 
observed at $b\lsim~+2^{\circ}$ (Figure~\ref{fig:f1}b).

\begin{figure}
\gridline{\fig{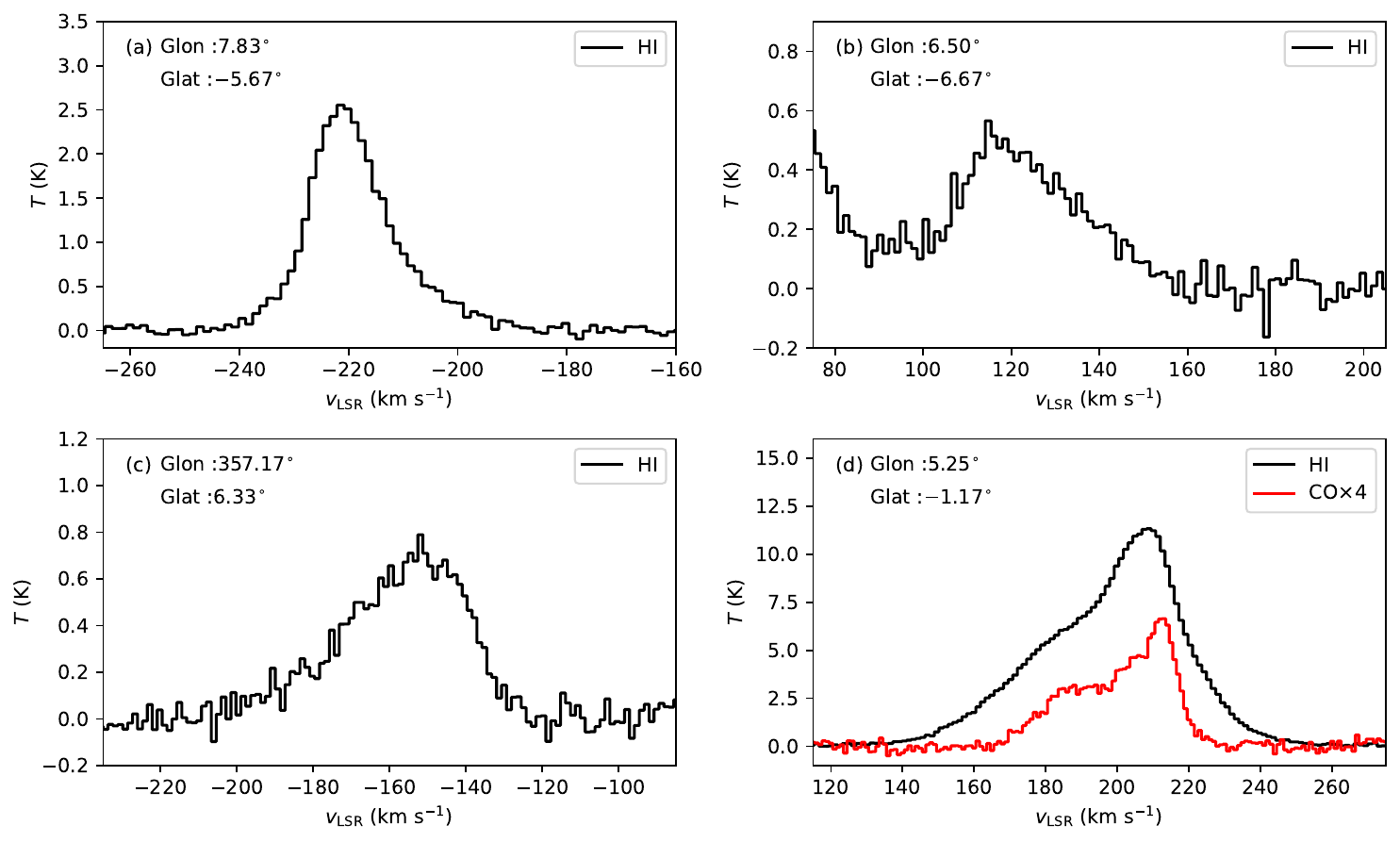}{1.0\textwidth}{} 
         }
\vspace{-6ex}
\caption{
Panels (a)--(c) show typical HVC spectral line profiles, while panel (d) displays
\HI\ and CO spectral profiles of the receding inflow gas from the near dust lane
\citep[e.g.,][]{2024ApJ...971L...6S,2025ApJ...984..109S}.
\label{fig:f3}}
\end{figure}

Furthermore, some HVCs in the near-side region display line profiles with sharp 
cutoffs at the blueshifted wing but extended broadening at the redshifted wing 
(Figure~\ref{fig:f3}(a) and (b)). These kinematic features differ 
from those of bar-driven gas inflows along the near dust lane, where both \HI\ 
and CO lines exhibit opposite spectral profiles (Figure~\ref{fig:f3}d). 
This evidence indicates that such clouds with negative velocities are 
collectively accelerating toward the observer, while the receding material with 
positive velocities 
in nearby regions is being systematically decelerated by certain mechanisms.
In the $l\lsim0^{\circ}$ region, the opposing spectral profiles imply that the 
approaching (i.e, blueshifted) flows on the far side are experiencing deceleration 
(Figure~\ref{fig:f3}c), while the receding (i.e., redshifted) streams show 
acceleration away from the GC. 

These kinematically anomalous HVCs along different lines of sight (LOSs) 
frequently exhibit broad line widths in both extreme negative-velocity components 
at $l\gsim7^{\circ}$ (FWHM$\sim17.9\pm8.1$~km~s$^{-1}$) and positive-velocity 
components at $l\lsim0^{\circ}$ (FWHM$\sim32.4\pm11.7$~km~s$^{-1}$).
The broadened line profiles of $\sim$20--30~km~s$^{-1}$ suggest strong 
turbulence and a high kinematic temperature of $\gsim7\times10^3$~K. This feature 
also shows that the approaching gas with narrower lines tends to lie closer to 
the observer (i.e., at the near side for $l\gsim0^{\circ}$ and $b\lsim0^{\circ}$) 
than the receding gas with broader ones 
(i.e., at the far side for $l\lsim0^{\circ}$ and $b\gsim0^{\circ}$).

In combination, the anomalous spatial and kinematic distributions of these 
HVCs, along with their large velocity dispersion and gradients, point to an 
origin involving extraordinary astrophysical processes. Meanwhile, the spectral 
and morphological properties distinguish them from the bar-driven inflows.
We thus suggest that a fraction of the cold gas located at considerable 
distances from the GP is perturbed by nuclear outflows, ultimately forming the 
observed HVCs. And we are characterizing the interaction interface, where 
multiphase outflows strip and entrain cold gas from the periphery of the tilted 
dust lanes and the inner warped gas disk, loading it into the Fermi/eROSITA bubbles. 

For example, Galactic winds have elevated clouds from the boundary of the 
near-side tilted dust lane, resulting in noticeable gas structures that extend 
vertically from the GP (i.e., considerable gas reaching $b\lsim-4^{\circ}$ or 
$z\lsim-500$~pc; see red contours near $l\sim+6^{\circ}$ in Figure~\ref{fig:f1}a).
The feature coexists with approaching gas with the extreme negative velocities 
\citep[blue contours near $l\sim+8^{\circ}$ in Figure~\ref{fig:f1}(a); 
also see][]{1974A&A....36..365S,2011A&A...533A.105W,2018MNRAS.474..289W}, 
which arises from overshooting streams that were accelerated toward the 
observer by the fast-moving winds. 
Indeed, the spectral profiles of both components align perfectly with 
the kinematic features of the outflowing gas mentioned above.

A notable asymmetry exists at the interface between the Fermi/eROSITA 
bubbles and the Galactic gaseous disk \citep[i.e., more CO in $z\lsim-260$~pc 
than in $z\gsim+260$~pc at $R_{\rm GC}\sim3$~kpc;][]{2022ApJ...930..112S}.
Interestingly, the disk-halo atomic clouds exhibit 
an excess in the comparable region of $l\sim17^{\circ}$--$35^{\circ}$
\citep[i.e., more \HI\ in $z\lsim-500$~pc;][]{2010ApJ...722..367F}.
The asymmetries agree with the idea of a higher mass-loading efficiency at 
negative $b$ for $l\gsim0^{\circ}$ (and at positive $b$ for $l\lsim0^{\circ}$).
We propose that the Galactic nuclear outflows primarily load cold gas from 
the tilted near (far) dust lanes and the distorted overshooting streams 
below (above) the GP.

The inclination of the Galactic bar (i.e., $\phi_{\rm bar}=23^{\circ}\pm3^{\circ}$)
imposes strong observational effects.
The LOS tangent plane of the southern X-ray bubble intersects the GP at a 
position slightly shifted northward \citep[extending beyond the GP rather 
than being rooted at the GC;][]{2024ApJ...967L..27L}, as its entrained gas 
from the near dust lane is preferentially below the GP. The gas entrainment 
within the northern bubble, influenced by the far dust lane, results in a
tangent plane that nearly aligns with the GP due to reduced projection effects.
The wind's interaction with the warped inner gas disk amplifies the overall 
geometric asymmetry of the eROSITA bubbles.

As an illustration, the Galactic outflows entrain gas from the outskirts of the tilted 
dust lanes, leading to an asymmetric distribution of material within the bubbles. 
In other words, the loaded gas on the near side is mainly concentrated
in the ($l\textgreater0^\circ$, $b\textless0^\circ$) region, while the far-side
part is predominantly at ($l\textless0^\circ$, $b\textgreater0^\circ$).
The enhanced gas loading promotes more efficient cooling ($t_{\rm cool}\propto n^{-1}$)
in the gas-rich interface between the tilted gas structures and the Galactic outflows.
We therefore propose that two factors may suppress X-ray emission: 
(1) strong radiative cooling rapidly dissipates the energy of shocks propagating 
through the higher-density environment (i.e., $n_{\rm cloud}\gsim0.1-1~{\rm {cm}^{-3}}$
in Section 4), and (2) the X-ray emission is subject to higher interstellar extinction 
near the GP, in addition to being attenuated by enhanced local absorption.
Our findings may provide a possible explanation for the observed X-ray asymmetries 
of the eROSITA bubbles. Namely, the X-ray emission is fainter in the interior of the
bubble and at the edges of the northwestern 
(i.e., far side, $\gsim$9~kpc) and southeast 
(i.e., near side, $\lsim$7~kpc) regions, particularly in the latter.

Cold and hot gas coexist within the eROSITA bubbles \citep[e.g.,][]
{2013ApJ...770L...4M,2016ApJ...826..215L,2018ApJ...855...33D,
2020Natur.584..364D,2020ApJ...888...51L,2025A&A...695A..60H,2026arXiv260107907D}. 
However, the presence of cold clouds, embedded in hot, high-velocity nuclear 
winds, challenges theoretical predictions of rapid cloud destruction. 
Traditional models anticipate cloud disruption via processes such as 
thermal, Kelvin-Helmholtz, and Rayleigh-Taylor instabilities, but 
struggle to account for the survival of these cold clouds over extended 
timescales \citep[e.g., see][]{2020A&ARv..28....2V}. 
To address this issue, researchers have proposed several mechanisms, 
including magnetic shielding, rapid cooling in turbulent radiative 
mixing layers (TRMLs), and the continuous replenishment of cold gas through 
thermal instability, extending the survival time of cold clouds 
in extreme environments \citep[e.g.,][]{2015ApJ...805..158S,
2018MNRAS.480L.111G,2020MNRAS.492.1970G,2020MNRAS.499.4261S,
2021MNRAS.502.3179T,2022MNRAS.510..551F,2024MNRAS.527.9683T}. 

Our results show that cold clouds entrained by Galactic outflows are 
preferentially located near gas-rich structures, such as tilted dust lanes, 
distorted overshooting streams, and warped inner gas disk. 
Consequently, these cold clouds have actually been subjected to disturbances 
for a much shorter time than previously assumed. This is logical, as these 
off-plane clouds were only recently swept up and entrained by the 
outflows rather than having undergone a long journey from the GC region.
Additionally, the fractal and porous structure of cold clouds enhances 
gas mixing and the production of a warm gas phase, facilitating rapid radiative 
cooling of the multiphase gas and effectively prolonging their survival 
time in such a gas-rich environment.

Observations via UV absorption-line spectroscopy have unambiguously 
confirmed the presence of substantial multiphase gaseous material within the 
eROSITA bubbles, exhibiting kinematic properties consistent with previously 
identified HVCs associated with nuclear winds \citep[e.g.,][]
{2015ApJ...799L...7F,2017ApJS..232...25S,2020ApJ...898..128A,2021ApJ...923L..11C,
2025ApJ...987L..32B}. The analyses also suggest that cool, low-ionization gas 
clouds ($\sim10^4$~K) are capable of surviving within the hot plasma of the 
bubbles, likely tracing fragmented clouds that have been carried outward by 
Galactic winds and outflows \citep[e.g.,][]{2017ApJ...834..191B,2023ApJ...944...65C}.
Recent evidence also reveals an atomic-to-molecular phase transition in the 
bubbles \citep[e.g.,][]{2023MNRAS.524.1258N}.
By integrating our finding that a considerable number of HVCs originate 
from tilted dust lanes and overshooting streams, this work largely reconciles 
the tension between the observed survival of cold clouds in the eROSITA
bubbles and theoretical predictions of their rapid destruction in extreme 
environments (i.e., high-velocity shocks and high temperatures and turbulence).

\section{Gas Circulation and Recycling in the Inner 3~kpc of the Galaxy}
\begin{figure}
\vspace{-5ex}
\gridline{\hspace{-5ex} \fig{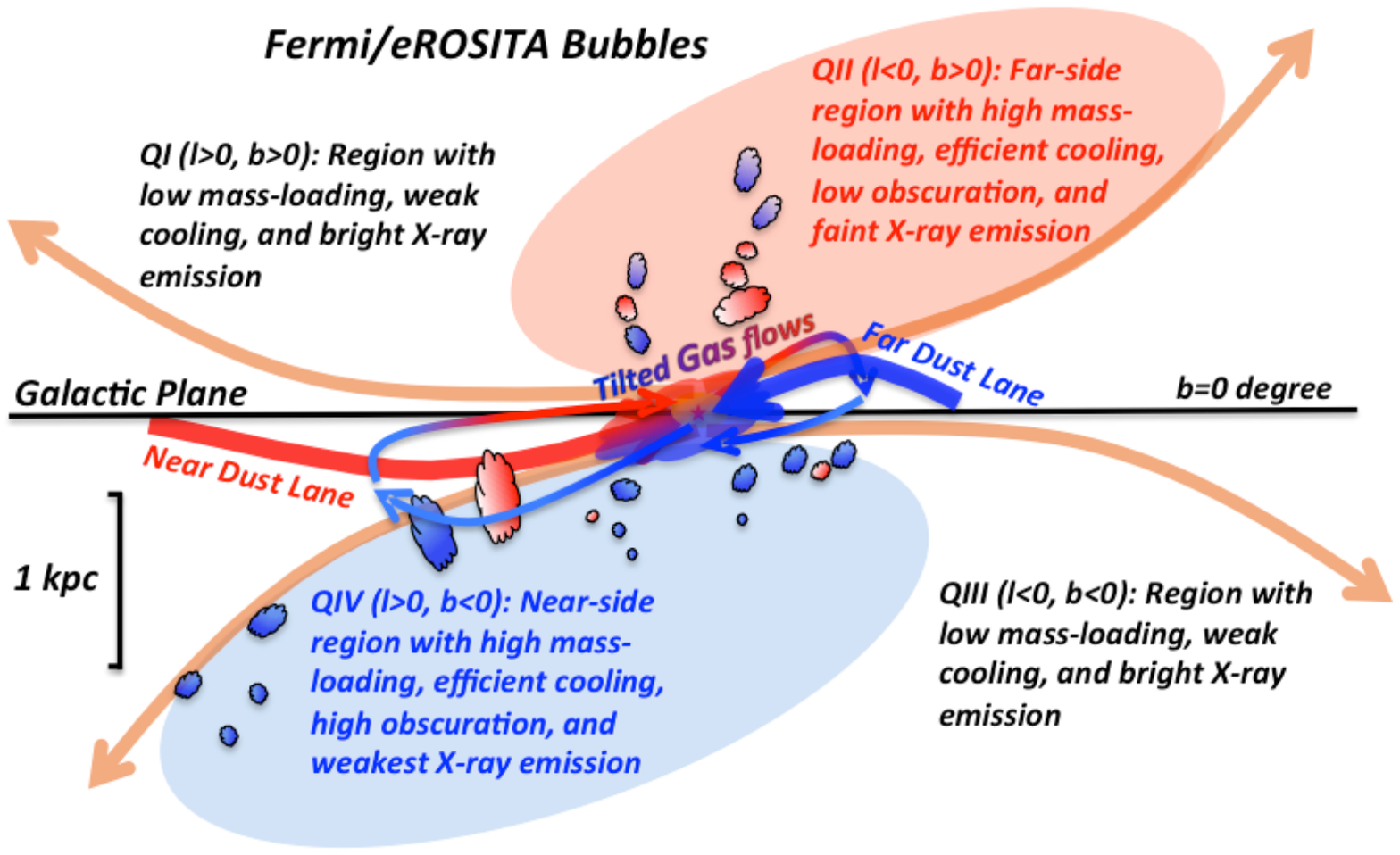}{1.0\textwidth}{} 
         }
\vspace{-10ex}
\caption{
Illustration of the gas circulation (i.e., inflows and outflows) toward the
inner Galactic region of $R_{\rm GC}\lsim3.0-3.4$~kpc. Red and blue arrows 
indicate receding and approaching gas flows toward the CMZ, respectively. 
The color blocks depict the fragmented HVCs driven by Galactic nuclear outflows. 
The four quadrants mark observational characteristics of multiphase gas in
term of mass-loading, cooling efficiency, and obscuration effects.
Galactic bar perturbations induce a globally tilted gas distribution, which 
in turn influences the large-scale symmetric patterns of the Galactic 
outflow structures in observations.
The orange solid arrows indicate the shock front, which corresponds to the 
boundary of the heavily obscured soft X-ray emission region near the base of 
the plane \citep[i.e., the possible spatial extent of the eROSITA bubbles at
lower Galactic latitudes of $|b|\lsim10^{\circ}$; refer to the X-ray and 
gamma-ray emission from the model of][]{2016ApJ...829....9M}.
\label{fig:f4}}
\end{figure}

Figure~\ref{fig:f4} illustrates a schematic view of gas circulation within
the inner 3~kpc of the Milky Way.
The model incorporates multiple components, including bar-driven inflows
(thick red and blue arrows), overshooting streams bypassing the outermost
region of the  CMZ (thin, twisted arrows), and clumpy clouds (colored blocks) 
entrained from the asymmetric gas structure into the Fermi/eROSITA bubbles.
The figure also marks variations in the Galactic biconical outflows across 
different regions, which are caused by cold gas loading from tilted and 
warped gas structures located 0.5--1.7~kpc from the GC. 

Overall, the mass loading is primarily modulated by Galactic outflow 
interactions with the tilted and warped gas layers in the QII 
(far, $\gsim$9~kpc) and QIV (near, $\lsim$7~kpc) regions.
On the contrary, QI and QIII, compared to QII and QIV, contain less 
entrained gas and therefore exhibit reduced dissipation, allowing them to better 
maintain their bubblelike morphologies in X-ray. On the other hand, dusty clouds 
near the GP largely obscure X-ray emission adjacent to the base of the Fermi/eROSITA 
bubbles, particularly in QIV where the near dust lane is situated. 

The gas-rich regions near the GP ($|z|\lsim0.4$~kpc or $|b|\lsim3^{\circ}$) 
are highly inhomogeneous, which promotes rapid cooling in porous clouds
that mix with the multiphase gas.
In this scenario, the global evolution of the Galactic outflows is 
governed by momentum exchanges with multiphase gas and regulated by energy 
dissipation through TRMLs and the related mechanisms 
\citep[e.g.,][]{2020ApJ...894L..24F,2021MNRAS.506.5658B,2022ApJ...924...82F,
2024MNRAS.530.4032C}.
These processes drive the essential thermal and kinematic coupling among 
the cold, warm, and hot gas phases, especially within the interaction layers 
between the Galactic outflows and the extraplanar gas structures.

In the region close to the CMZ, the nuclear winds are likely confined by the
high gas density and pressure. Upon achieving breakout from the dense CMZ 
region \citep[characterized by radial extent $R\sim$200--300~pc for molecular 
gas;][]{2023ASPC..534...83H}, the powerful nuclear winds undergo rapid lateral 
expansion. That is, beyond the maximum vertical extent\footnote{Toward the 
inner Galaxy, the thick molecular disk is approximately 3 times thicker 
than the thin disk \citep[280~pc vs. 85~pc; see Table~3 in][]{2021ApJ...910..131S}. 
Assuming that the maximum vertical extent of gas in the CMZ is broadly 
characterized by the thick molecular component, we derive a typical scale 
height of $H_{\rm z}\sim 60$~pc, which is comparable to that of the atomic gas 
with a Gaussian vertical 
distribution of $n({\rm H})=n_0\times$exp(-$\frac{z^2}{H_{\rm z}^2}$) 
\citep[][]{2007A&A...467..611F}.} of the CMZ (e.g., twice the scale 
height of $\sim2\times H_{\rm z}\sim$120~pc or $|b|\lsim 0\fdg9$ at 8.15~kpc), 
the winds expand radially, developing outflows with a wider opening angle 
of $\gsim140^{\circ}$ (i.e., based on the tilted angle of the gas flows 
of $\theta_{\rm gas\ flows}\lsim 20^{\circ}$; Section 2).

As outflows propagate through the inner Galaxy, they interact with the warped 
disk, lifting up clouds from these preexisting, extraplanar structures
at $R_{\rm GC}\sim0.5-1.7$~kpc.
Consequently, the outflows push the clouds outward to larger $R_{\rm GC}$ and 
upward to higher altitudes (e.g., $|z|\gsim$250~pc or $|b|\gsim 2\fdg0$), 
guiding the multiphase gas in multiple directions. 
The Galactic outflows preferentially target cold clouds along the peripheries
of the tilted dust lanes, where they strip, entrain, and continuously load 
a bulk of gas into the Fermi/eROSITA bubbles. By acting on the gas-rich interfacial 
layers, the Galactic outflows alter their own propagation dynamics, regulate 
large-scale vertical gas circulation, and ultimately shape the bubbles' 
multiwavelength properties (Figure~\ref{fig:f4}).

If the HVCs are weakly coupled to Galactic rotation and have a wide opening 
angle ($\gsim140^{\circ}$), we can neglect their rotation and projection 
effects relative to the GP (i.e., low $|b|$). The outflow velocity can be 
approximated as
\begin{equation}
v_{\rm {outflow}}=\frac{v_{\rm {LOS}}+\Omega_{\rm {Sun}}\times{R_{\rm {sun}}\times{\rm {sin}}(l)}}{{\rm {cos}}(\phi_{\rm {HVC}}+l)}
\end{equation}

Here, $v_{\rm {outflow}}$ represents the outflow velocity, measured using 
HVCs that are entrained by high-velocity outflows driven
by the Galactic nuclear winds,
and $v_{\rm {LOS}}$ is the observed LSR velocity of the moving gas along the LOS.
$l$ and $\phi_{\rm {HVC}}$ are the longitude and the inclination angle of the HVC, 
respectively, and $\Omega_{\rm {sun}}$ and $R_{\rm {Sun}}$ are the rotation speed 
and the Galactocentric distance of the Sun, respectively \citep[$\Omega_{\rm {Sun}}
=30.32\pm0.27~\km\ps\kpc^{-1}$ and $R_{\rm {Sun}}=8.15\pm0.15$~kpc;][]{Reid19}.

Considering that a high-velocity outflow breaks through the CMZ and loads cold 
gas with a warped gas distribution ($\phi_{\rm {HVC}}\sim1.5\times\phi_{\rm {bar}}$; 
Section 2) into the outflows, 
we find that the outflow velocity traced by HVCs at the near side 
(Figure~\ref{fig:f2}a for the QIV region) is $\sim-230~\km\ps$ at $l\sim9^{\circ}$ 
and $\sim-340~\km\ps$ at $l\sim23^{\circ}$, respectively. 
The estimated velocities of entrained gas agree well with relatively 
mild shocks in the 2.3$\times10^6$~K warm-hot gas \citep[i.e., Mach number of 
$\sim$1--1.5 from $\sim0.2$~keV X-ray emission at the bubble's bright edge;
see the eROSITA and Suzaku observations from][]
{2020Natur.588..227P,2023NatAs...7..799G}. 

It is worth noting that the structure of the eROSITA bubbles is not clearly 
visible at low Galactic latitudes \citep[e.g.,][]{2024ApJ...967L..27L}. 
Confirming the presence of $\sim 10^6$~K thermal gas in such regions 
(e.g., $|b|\lsim10^{\circ}$ or $|z|\lsim$~1~kpc) remains a significant 
challenge because of the strong absorption of soft X-rays by the abundant 
interstellar dust and gas near the GP. 
Despite this, a substantial number of HVCs are sandwiched between the
Fermi bubble's inner region (hot, shocked wind) and its outer shell structures 
\citep[shocked ISM and CGM; see QIV's HVCs outside the orange dotted lines in 
Figure~\ref{fig:f1}; also refer to the models in][]
{2015MNRAS.453.3827S,2017MNRAS.467.3544S,2016ApJ...829....9M}.

In the multiphase nuclear outflow, cold clouds are not merely passive 
tracers but coevolving components dynamically coupled to the warm--hot wind. 
The entrained HVCs are influenced by a combination of ram pressure from 
the outflows and, more critically, continuous mass, momentum, and energy 
exchange mediated by TRMLs \citep[][]{2020ApJ...894L..24F,2022ApJ...924...82F} 
at the interface between the expanding winds and the gas-rich regions. 
This exchange regulates cloud survival, enabling clouds to gain or lose mass 
while being accelerated, and fundamentally modifies the outflow's structure 
compared to single-phase models.

Therefore, these HVCs likely trace gas in the interface layer between the 
expanding bubbles (i.e., the inner Fermi bubbles) and the inhomogeneous, 
warped gaseous disk (i.e., the outer eROSITA bubbles near the GP). 
The apparent locations of HVCs inside the northern Fermi bubbles are primarily 
a projection effect. We argue that the HVCs, which manifest as a shocked 
and rapidly cooling shell, represent the cold embedded component of the
dynamically coupled Galactic outflows that is interacting with the warped 
gaseous disk. The physics of TRMLs governs HVCs' entrainment and evolution 
in multiphase Galactic outflows.
Recently, many high-resolution simulations have explored these intriguing 
kinematic and multix-wavelength features \citep[e.g.,][]{2015MNRAS.453.3827S,
2017MNRAS.467.3544S,2021MNRAS.506.5658B,2022MNRAS.511..859G,2026MNRAS.545f2065W}.

In QIV, the two extreme-velocity HVC populations each occupy at least 
$\sim10\%$ of the projection area, indicating a considerable spatial filling 
factor for the entrained gas. Due to significant contamination from unrelated 
gas emission, it is difficult to unambiguously identify other perturbed gas 
components within the velocity range of $|V_{\rm LSR}|\lsim100~\km\ps$ 
that are associated with the Galactic nuclear outflows.
Nevertheless, leveraging current observations, we can obtain meaningful limits 
on the total perturbed gas mass affected by the large-scale outflows. 

For the near-side region of $l\gsim+5^{\circ}$, assuming that the LOS
thickness of the identified HVCs is comparable to their projected size on the sky,
we can estimate their typical volume density, i.e., $\sim0.2~{\rm {cm}^{-3}}$
for the negative-velocity component (blue contours for receding gas in
Figure~\ref{fig:f1}a) and $\sim1.0~{\rm {cm}^{-3}}$ for the positive-velocity
component (red contours for gas decelerated and lifted by outflows).
In total, the cumulative mass of these observed HVCs in QIV is
$\gsim0.1\times10^6~\Msun$.

In fact, considering the complex velocity distribution of the clouds toward 
the GC, the HVCs identified in QIV likely represent only a small fraction of 
the total gas entrained by the Galactic nuclear outflows. 
Under the assumption of a roughly flat gas velocity distribution,  
the observed HVCs would only cover about 20\%--30\% of the full velocity range.
The actual mass of gas entrained by the Galactic outflows would be a factor of 
$\sim$5 greater than the directly observed value. Therefore, within the QIV region 
(i.e., $\sim$2~kpc long $\times$1~kpc high $\times$0.5~kpc thick), 
the total mass of the neutral gas driven by Galactic outflows can reach 
$\gsim0.5\times10^6~\Msun$.  

Given that cold gas is entrained from the Galactic inner warped disk into 
the bubble's cavity by nuclear outflows
(i.e., from $R_{\rm GC}\sim$~0.5--1.0~kpc to $R_{\rm GC}\sim$~1.7--3.4~kpc), 
a travel distance of $\sim$1--2~kpc corresponds to a dynamical timescale
of $\sim3$--6~Myr. Consequently, the loading rate of the cold atomic gas 
in QIV is conservatively estimated to be
$\gsim0.5\times10^6~\Msun$/(3--6~Myr)$=0.1$--$0.2~\Msun\yr^{-1}$.
Incorporating the gas from both the tilted gas structures (i.e., near and far dust
lanes and overshooting streams in QII and QIV, $\gsim0.2$--$0.4~\Msun\yr^{-1}$) 
and the inner dense regions near the CMZ \citep[$\sim0.1$~$\Msun\yr^{-1}$; e.g.,]
[]{2003ApJ...582..246B,2019Natur.567..347P,2021A&A...646A..66P,
2023A&A...674L..15V}, the cold gas loading rate may exceed 
0.3--$0.5~\Msun\yr^{-1}$ for the biconical outflows. Furthermore, accounting 
for the ionization of neutral gas by the energetic winds 
\citep[e.g.,][]{2017ApJ...834..191B,2019ApJ...886...45B,2023ApJ...944...65C,
2023ApJ...954...64S} and an ionization fraction of $\sim$55\% 
\citep[][]{2020SciA....6.9711K}, the total mass-loading rate of Galactic 
outflows could reach values $\gsim0.6-1~\Msun\yr^{-1}$, roughly matching 
the magnitude to the inflow rate induced by the Galactic bar 
\citep[e.g.,][]{2024ApJ...971L...6S,2025ApJ...984..109S}.

So far, we know that gas dynamics are crucial for governing the gas distribution 
and circulation within the inner Galaxy. Taken together, gas inflows are funneled 
toward the CMZ by the Galactic bar. During the dynamical processes, the gas flows 
develop tilted and warped structures relative to the GP. Simulations show that the 
gas streams trace the Galactic bar's nonstationary nature, producing large-amplitude 
oscillations and an off-centered gas distribution in the inner 3~kpc of the Galaxy 
\citep[e.g.,][]{1999A&A...345..787F,2008A&A...477L..21M,2025PASA...42...14C}.
At the same time, the giant Fermi/eROSITA bubbles, blown by episodic energy 
release from the GC, significantly modify both the kinematic and spatial properties 
of gas through interactions between the Galactic nuclear outflows and the 
tilted and warped gas structures.
By highlighting that the nuclear winds load cold gas from these preexisting
gas-rich structures into the bubbles, our study offers a new perspective on 
how large-scale structures effectively regulate gas circulation in the inner Galaxy 
and influence long-term Galactic evolution.

\section{Summary}
The impact of galaxy-scale nuclear winds and outflows on environmental feedback has been 
extensively examined through numerical simulations \citep[e.g.,][]{1999ApJ...513..142M,
2008ApJ...674..157C,2015MNRAS.453.3827S,2017MNRAS.466.3810F}.
High-velocity gas, entrained and lifted by large-scale galactic outflows, has also
been widely observed in nearby galaxies such as M82, NGC~253, and NGC~3079 
\citep[e.g.,][]{2020A&ARv..28....2V,2024A&ARv..32....1S}.

In the Milky Way, we show that the anomalous spatial distribution and 
kinematics of the HVCs are attributable to large-scale Galactic outflows with 
a full opening angle of $\gsim140^{\circ}$, which load and entrain off-plane cold 
gas to higher latitudes. The farther the cloud is from the GC, the greater its velocity. 
During the millions of years of gas acceleration, the entrained clouds can reach 
velocities $\gsim300~\km\ps$ while experiencing progressive disintegration and 
dissipation. We propose, for the first time, that the cold clouds within the 
Fermi/eROSITA bubbles originate not exclusively from regions near the GC, 
but substantially from outer regions at $R_{\rm GC}\sim$0.5--1.7~kpc, 
where the nuclear winds break through the CMZ and interact with the preexisting 
warped structures (e.g., tilted dust lanes and distorted overshooting streams) 
governed by the Galactic bar and bulge.

In fact, the tilted dust lanes, coupled with distorted overshooting streams,
not only promote gas inflows into the CMZ but also provide critical pathways
for Galactic outflows to strip, entrain, and load gas from these extraplanar
structures into the Fermi/eROSITA bubbles. The nuclear-wind-dominated outflows 
and bar-driven inflows roughly achieve global equilibrium, operating at a
comparable rate on the order of $\sim1~\Msun$~yr$^{-1}$.
Given the relatively low star formation rate ($\sim0.1~\Msun$~yr$^{-1}$) over
the past tens of millions of years in the CMZ and the inferred high gas outflow
rate, therein, the origin of the large-scale bubbles likely stems from the 
long-term accumulation of episodic starbursts in the CMZ and intermittent 
accretion and jet processes associated with the SMBH at the GC.

The tilted dust lanes, as well as the overshooting effect, regulate
the spatial distribution, physical properties, and dynamical evolution of the
gas in the inner Galactic disk, while they also complicate the spatial and
velocity distribution of the entrained gas within the large-scale bubbles.
The inflows consist primarily of molecular gas and remain confined to the
Galactic gas disk. In contrast, the nuclear-wind-driven outflow are multiphase,
propagating along the interface between the outflowing material and
warped gas structures into the bubbles.

In this scenario, the clumpy, porous clouds entrained in the nuclear outflows 
facilitate radiative cooling through TRMLs, thereby enhancing cold gas mass 
loading at interacting layers of multiphase gas. 
And asymmetric gas entrainment provides a possible explanation for 
both the observed morphological asymmetry of the eROSITA bubbles and their 
asymmetric interaction with the gaseous disk at $R_{\rm GC}\sim$~3.0--3.4~kpc.
These new findings necessitate revisions to models of Galactic nuclear outflows, 
as simulations systematically underestimate the persistence and mass-loading 
capacity of cold gas in the outflows by neglecting the large-scale tilted 
and warped gas structures in the inner Galaxy.

\begin{acknowledgments}
We are grateful to the referee for the careful reading and insightful
comments, which have significantly improved the quality of the manuscript.
This research made use of the data from the Milky Way Imaging Scroll Painting
(MWISP) project, which is a multiline survey of \twCO,\thCO,C$^{18}$O along the
northern Galactic plane with the PMO-13.7m telescope. We are grateful to all the 
members of the MWISP working group, particularly the staff members at the PMO-13.7~m 
telescope, for their long-term support. MWISP was sponsored by National Key R\&D 
Program of China with grants 2023YFA1608000 \& 2017YFA0402701 and by CAS Key Research 
Program of Frontier Sciences with grant QYZDJ-SSW-SLH047.
Y.S. acknowledges funding from the Basic Research Program of Jiangsu 
(BK20252109) and the National Natural Science Foundation of China (12173090).
The work makes use of publicly released data from the HI4PI survey, which combines
the EBHIS in the Northern Hemisphere with the GASS in the Southern Hemisphere.
The Parkes Radio Telescope is part of the Australia Telescope National Facility,
which is funded by the Australian Government for operation as a National Facility
managed by CSIRO. The EBHIS data are based on observations performed with the
100 m telescope of the MPIfR at Effelsberg. EBHIS was funded by the Deutsche
Forschungsgemeinschaft (DFG) under the grants KE757/7-1 to 7-3.
\end{acknowledgments}

\facility{PMO:DLH}

\bibliography{references}{}
\bibliographystyle{aasjournal}

\clearpage

\clearpage

\clearpage

\clearpage

\end{document}